\documentclass[12pt,a4paper]{article}
\RequirePackage{lineno}
\usepackage{epsfig}
\usepackage{graphicx}                  
\usepackage{siunitx}
\usepackage{ae}
\usepackage{amsmath}
\usepackage{amssymb}
\usepackage{graphics}
\usepackage{dcolumn}
\usepackage{bm}
\usepackage[symbol]{footmisc}
\usepackage{setspace}
\usepackage{ragged2e}
\usepackage{wrapfig}
\begin{document}

\centering

{\Large
			{\bf Long term stability study of triple GEM detector using different Argon based gas mixtures: an update}}
\date{}


\centering
{S Chatterjee\footnote[1]{Corresponding Author:
sayakchatterjee896@gmail.com, sayakchatterjee@jcbose.ac.in}, S Roy, A Sen, S Chakraborty\footnote[2]{Present address:
Variable Energy Cyclotron Centre, Kolkata, India}, S Das, \\ S K Ghosh, S K Prasad, S Raha and S Biswas}

\vspace*{0.5cm}
{Department of Physics and Centre for Astroparticle Physics and Space Science (CAPSS), ~Bose Institute, EN-80, Sector V, Kolkata 700091, India}

\vspace*{0.5cm}
\centering{\bf Abstract}
\justify

The long-term stability in terms of gain and energy resolution of a prototype triple Gas Electron Multiplier~(GEM) detector has been investigated with high rate X-ray irradiation. Premixed Ar/CO$_{2}$ (80:20) and (90:10) gases have been used for this stability study. A strong Fe$^{55}$ X-ray source is used to irradiate the chamber. The uniqueness of this work is that the same source has been used to irradiate the GEM prototype and also to monitor the spectra. This arrangement is important since it reduces the mechanical complexity of using an X-ray generator as well as the cost of the setup. A small area of the chamber is exposed continuously to the X-ray for the entire duration of the operation. The effect of temperature and pressure on the gain and energy resolution is monitored. The result of the long-term stability test for a triple GEM detector using Ar/CO$_{2}$ (70:30) gas mixture has been reported earlier~\cite{1}. The results with Ar/CO$_{2}$ (80:20) and (90:10) gas mixtures for the same chamber are presented in this article. 

\section{Introduction}
Gas Electron Multiplier detectors are being used in many High Energy Physics experiments as tracking devices for their high rate capability~\cite{2}\cite{3}\cite{4}.~The Compressed Baryonic Matter~(CBM)~\cite{5} experiment at the future Facility for Antiproton and Ion Research (FAIR)~\cite{6}, Darmstadt, Germany will use the triple GEM detector as a tracking device in the MUon CHamber~(MUCH)~\cite{7}\cite{8}\cite{9}\cite{10} to track the di-muon pairs originating from the decay of low mass vector mesons, which will give us the information about the fireball created after the collision. Triple GEM detectors will be used in the first two stations of MUCH because the rate will be very high~($\sim$1.0 MHz/cm$^2$ for the first station~\cite{11}) there. The motivation of this particular work is to understand how the chamber behaves under continuous high irradiation~\cite{1}\cite{12}\cite{13}. The stability study in terms of gain and energy resolution of a triple GEM detector has been carried out with Ar/CO$_{2}$ (80:20) and (90:10) gas mixtures and results are presented in this paper.      
\section{Experimental details}											
A double mask triple GEM detector prototype obtained from CERN, having dimensions 10~cm$\times$10~cm,  has been used for this study. The drift, transfer and induction gaps of the detector are kept at 3~mm, 2~mm, 2~mm, respectively. The high voltage is distributed between the drift plane and individual GEM foils using a voltage divider resistor chain. The signal is collected using a sum up board from all the segmented readout pads, each of dimensions 9~mm~$\times$~9~mm. It is fed to a charge sensitive preamplifier~(VV50-2) having a gain of 2~mV/fC and shaping time of 300 ns~\cite{14}. The output of the preamplifier is fed to the linear Fan In Fan Out~(FIFO) module. One output of the linear FIFO is fed to a Single Channel Analyzer~(SCA) for rate measurement and the other output is fed to a Multi-Channel Analyzer~(MCA) to obtain the energy spectrum. The SCA is operated in integral mode and the threshold is set at 0.1 V to reject noise. The output of the SCA is fed to the TTL-NIM adapter to convert the TTL signal to NIM signal and then counted by a NIM scaler. \\
Premixed Ar/CO$_{2}$ (80:20) and (90:10) gases have been used at a constant flow rate for this study. $\Delta$V$\sim$ 359~V and 331~V across each GEM foil were maintained throughout the experiment for (80:20) and (90:10) gas mixtures, respectively. A perspex collimator having an area of $\sim$13~mm$^2$ has been used to irradiate the chamber at a rate of $\sim$250 kHz. The results obtained for the (80:20) and (90:10) gas mixture are given in the next section.         
\section{Results}
The gain and energy resolution have been calculated from the $Fe^{55}$ 5.9 keV peak and then normalization is done using T/p correction as reported earlier~\cite{1}. The stability test of the detector has been carried out for $\sim$30~hours and $\sim$140~hours with Ar/CO$_2$ (80:20) and (90:10) gas mixtures, respectively.

\begin{figure}[htbp]
	\centering
		\includegraphics[scale=0.50]{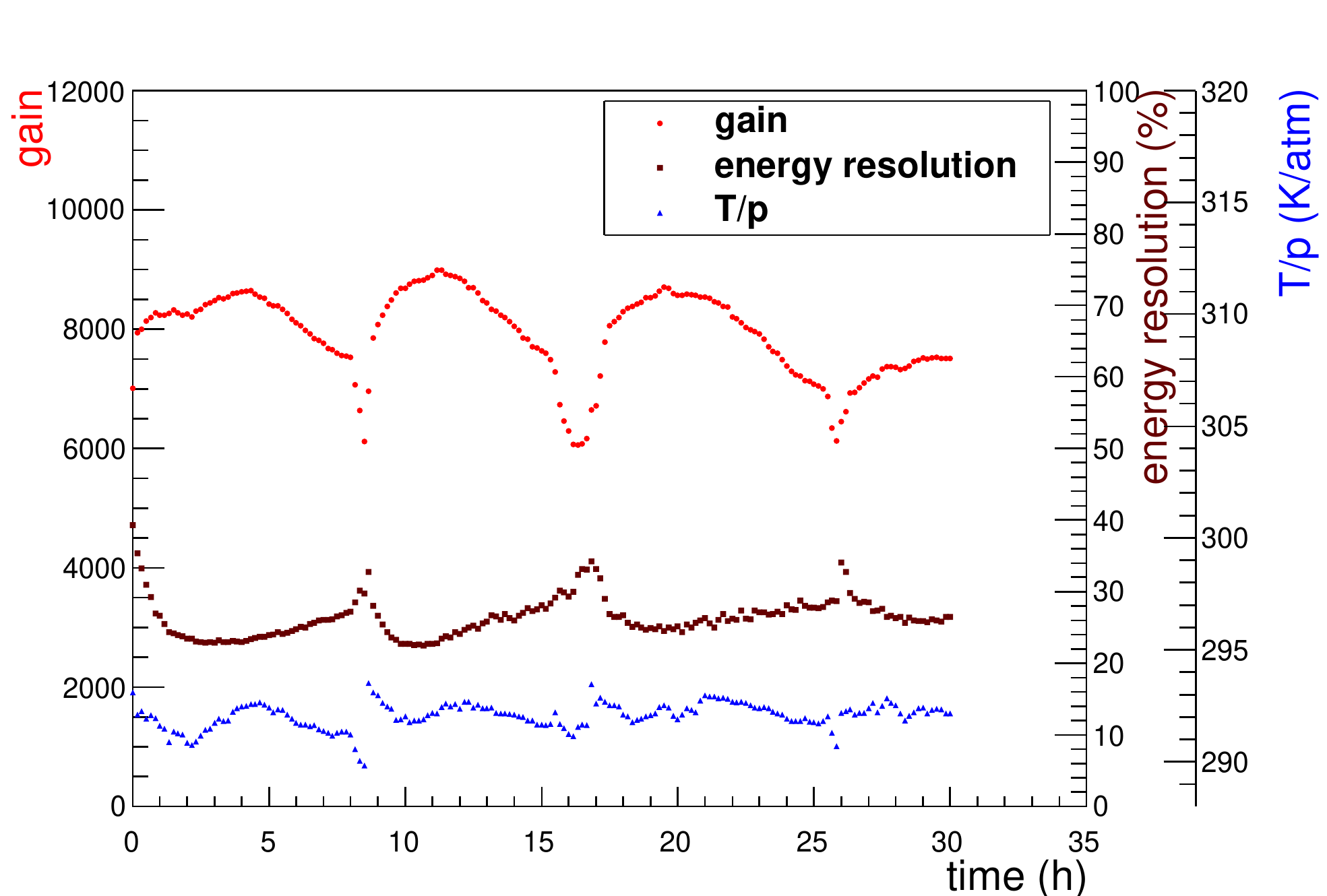}
		\caption{\label{label}Variation of gain, energy resolution and T/p as a function of time for Ar/CO$_2$ (80:20) gas mixture. }
\end{figure}		
\begin{figure}[htbp]
	\centering
		\includegraphics[scale=0.50]{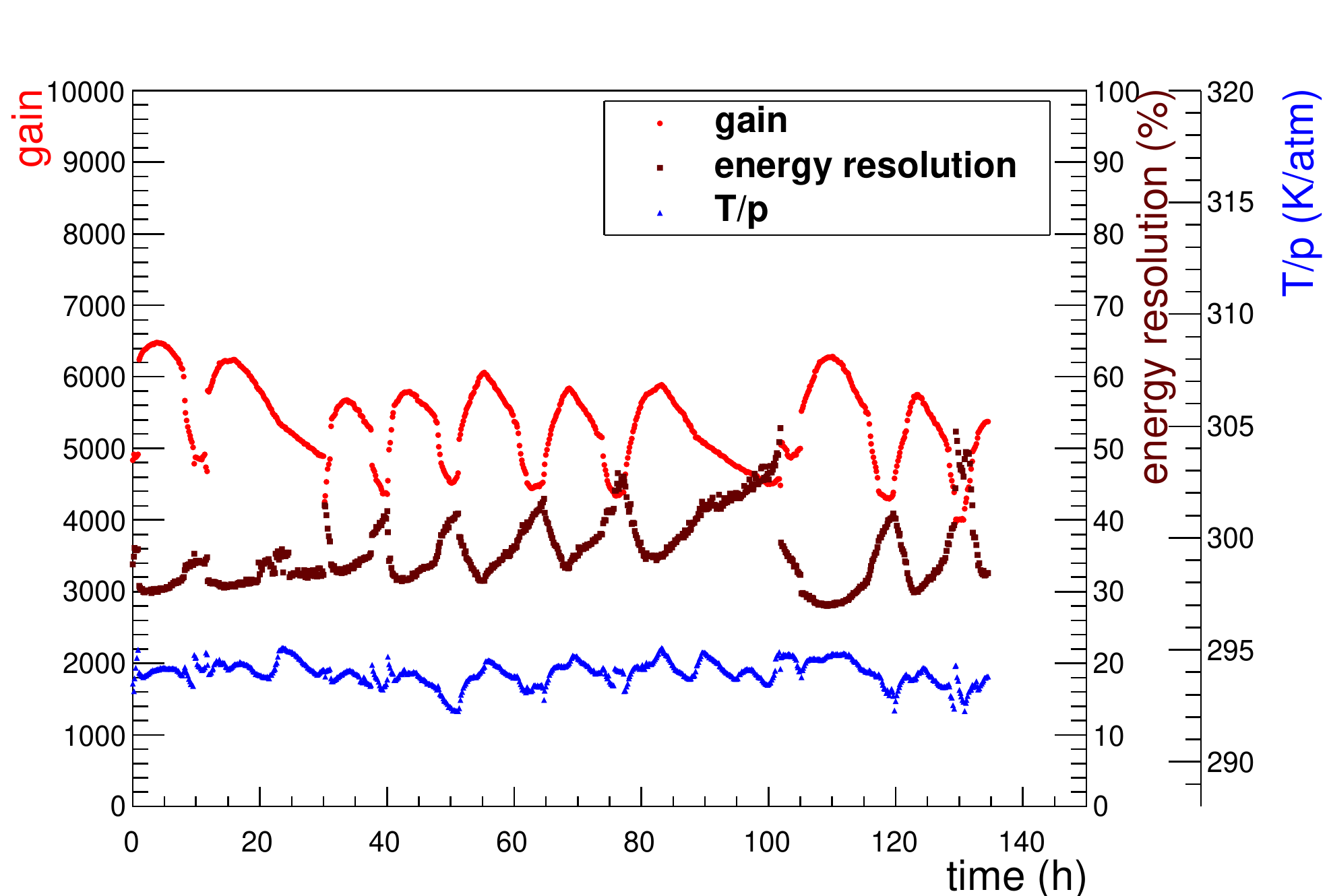}
		\caption{\label{label}Variation of gain, energy resolution, and T/p as a function of time for Ar/CO$_2$ (90:10) gas mixture.}
\end{figure}

\begin{figure}[htbp]
	\centering
		\includegraphics[scale=0.50]{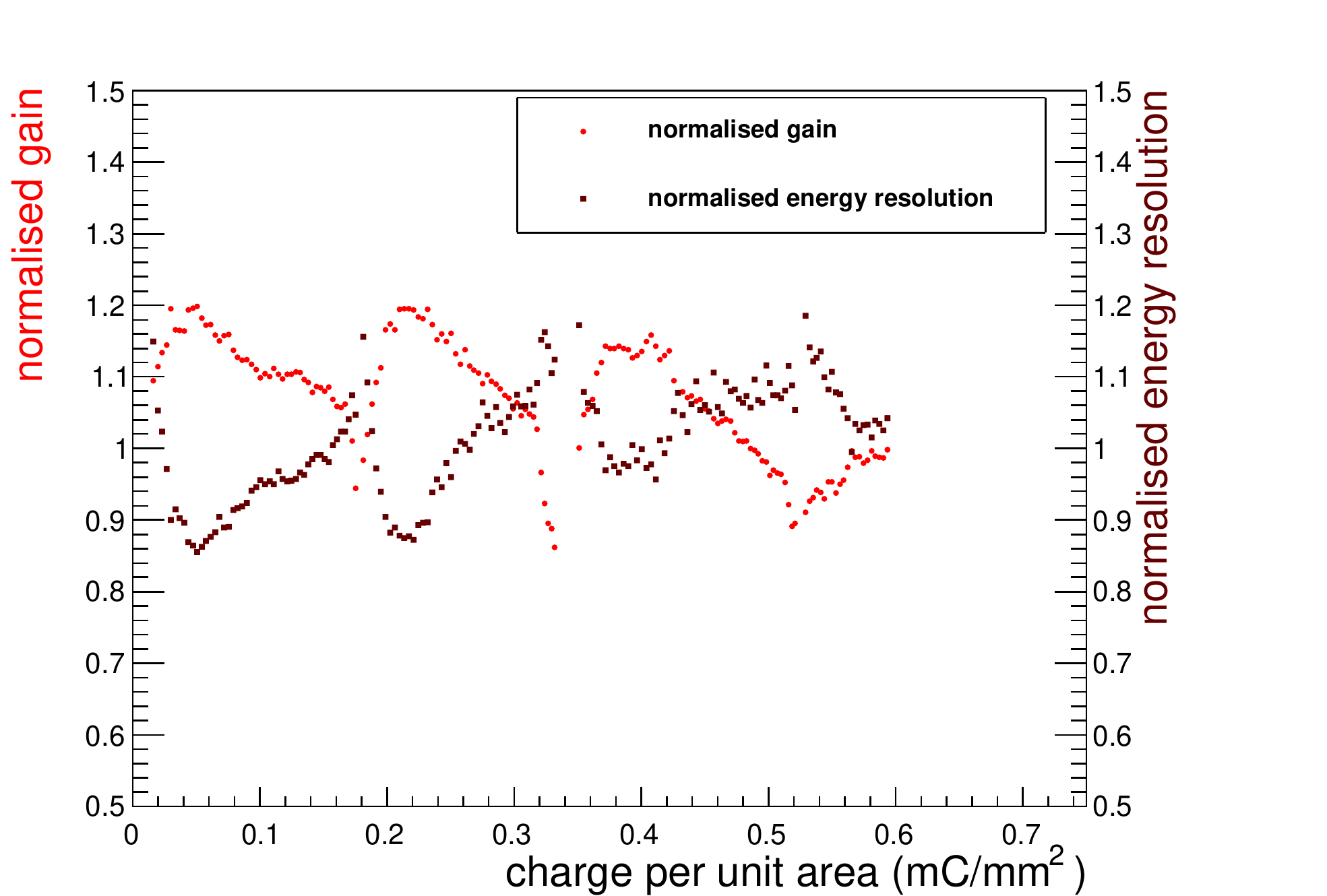}
		\caption{\label{label}Variation of normalized gain and normalized energy resolution as a function of the total charge accumulated per unit area for Ar/CO$_2$ (80:20) gas mixture}
	\end{figure}	

\begin{figure}[htbp]
	\centering
		\includegraphics[scale=0.50]{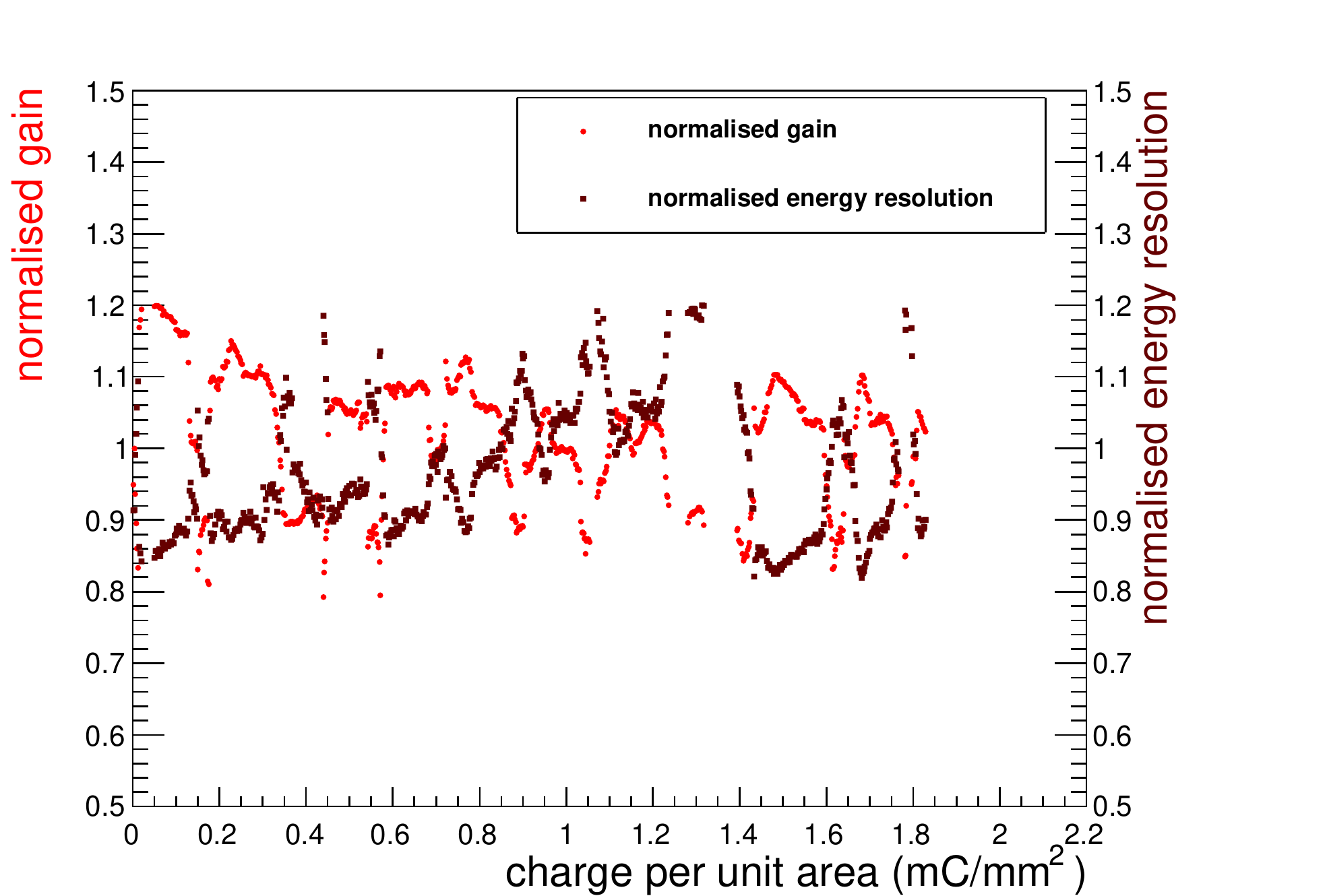}
		\caption{\label{label}Variation of normalized gain and normalized energy resolution as a function of the total charge accumulated per unit area for Ar/CO$_2$ (90:10) gas mixture} 
\end{figure}

\begin{figure}[htbp]
	\centering
		\includegraphics[scale=0.50]{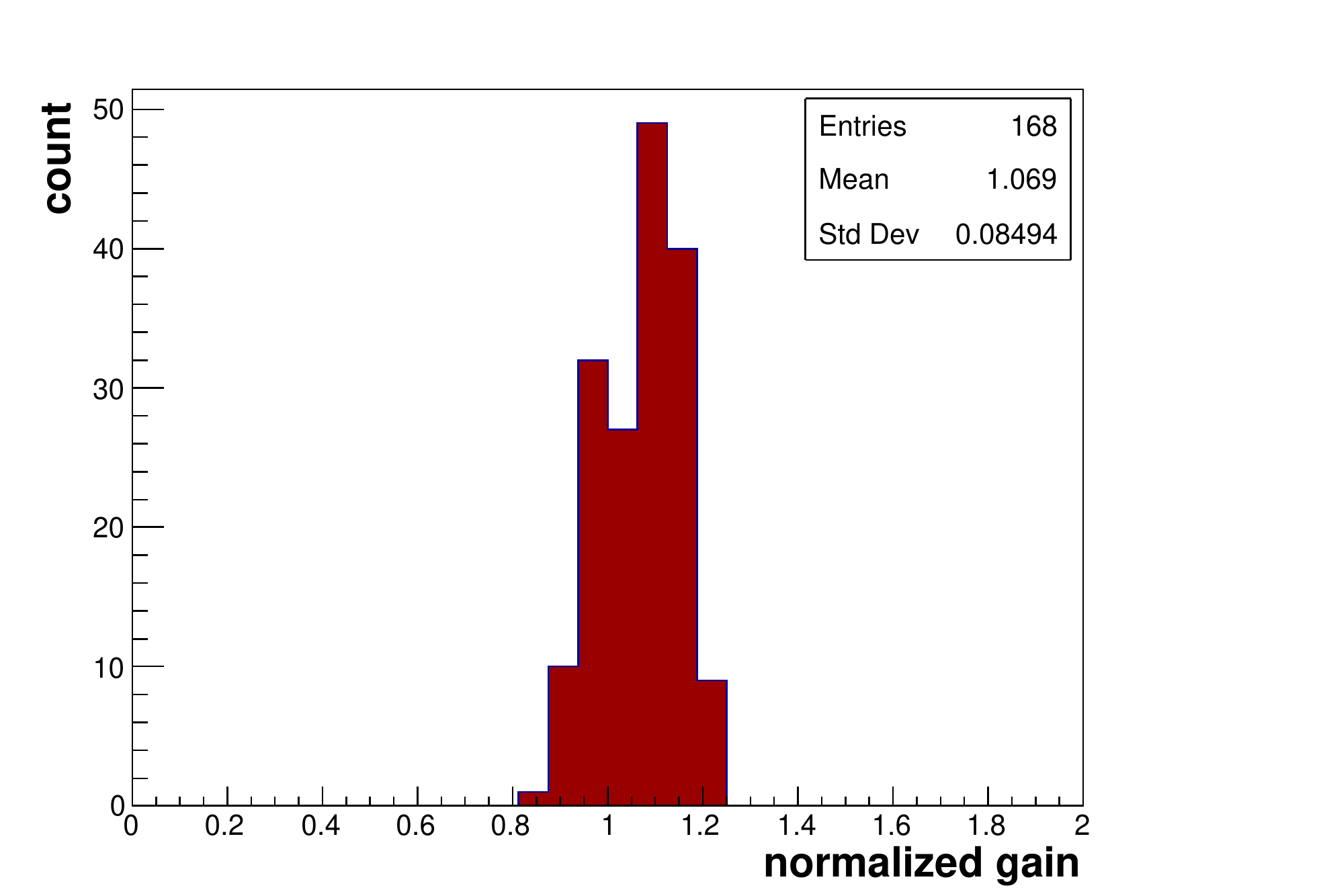}
		\caption{\label{label}Distribution of normalized gain for Ar/CO$_2$ (80:20) gas mixture}
\end{figure}
\begin{figure}[htbp]
	\centering	
		\includegraphics[scale=0.50]{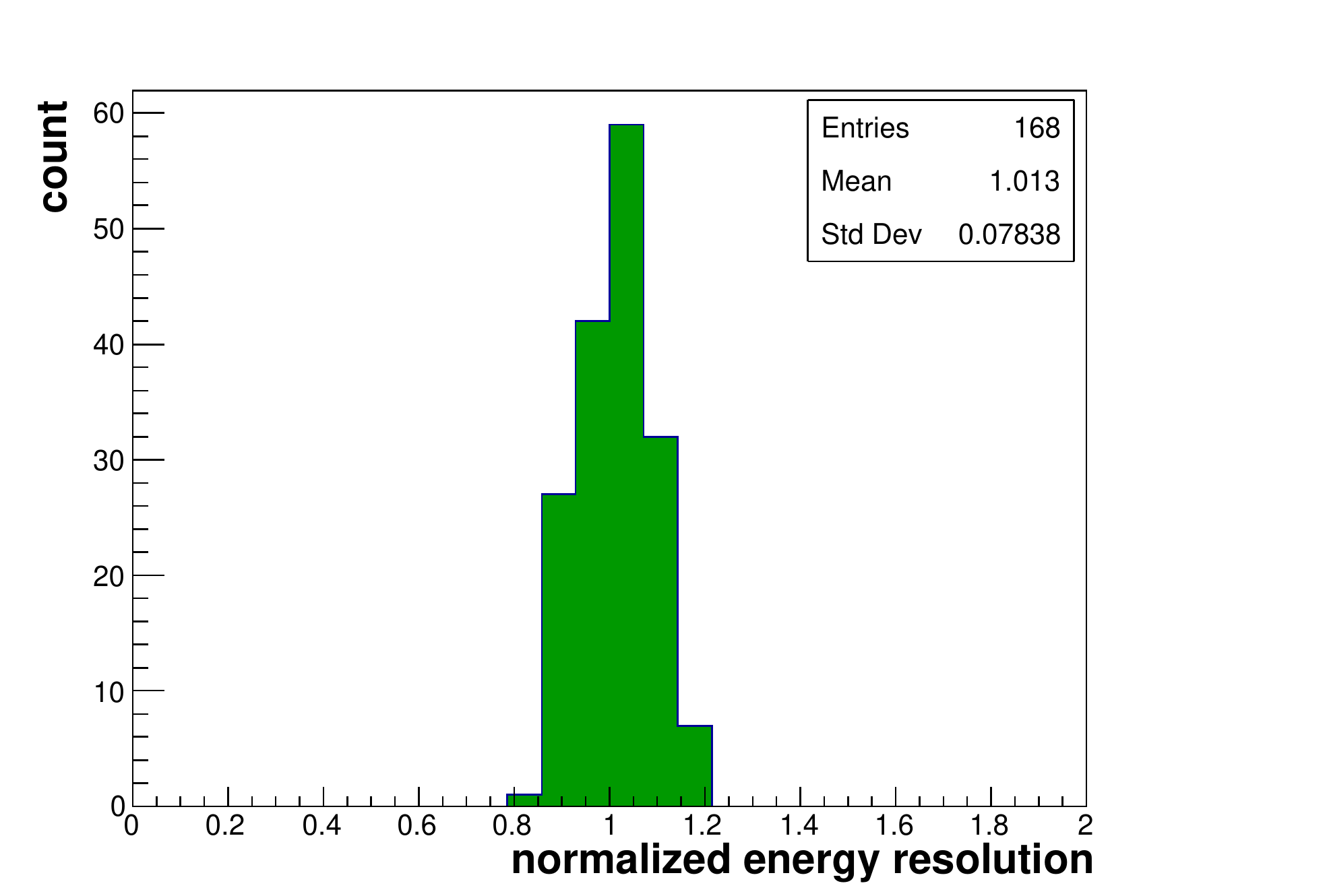}
		\caption{\label{label}Distribution of normalized energy resolution for Ar/CO$_2$ (80:20) gas mixture} 
\end{figure}

\begin{figure}[htbp]
	\centering
		\includegraphics[scale=0.50]{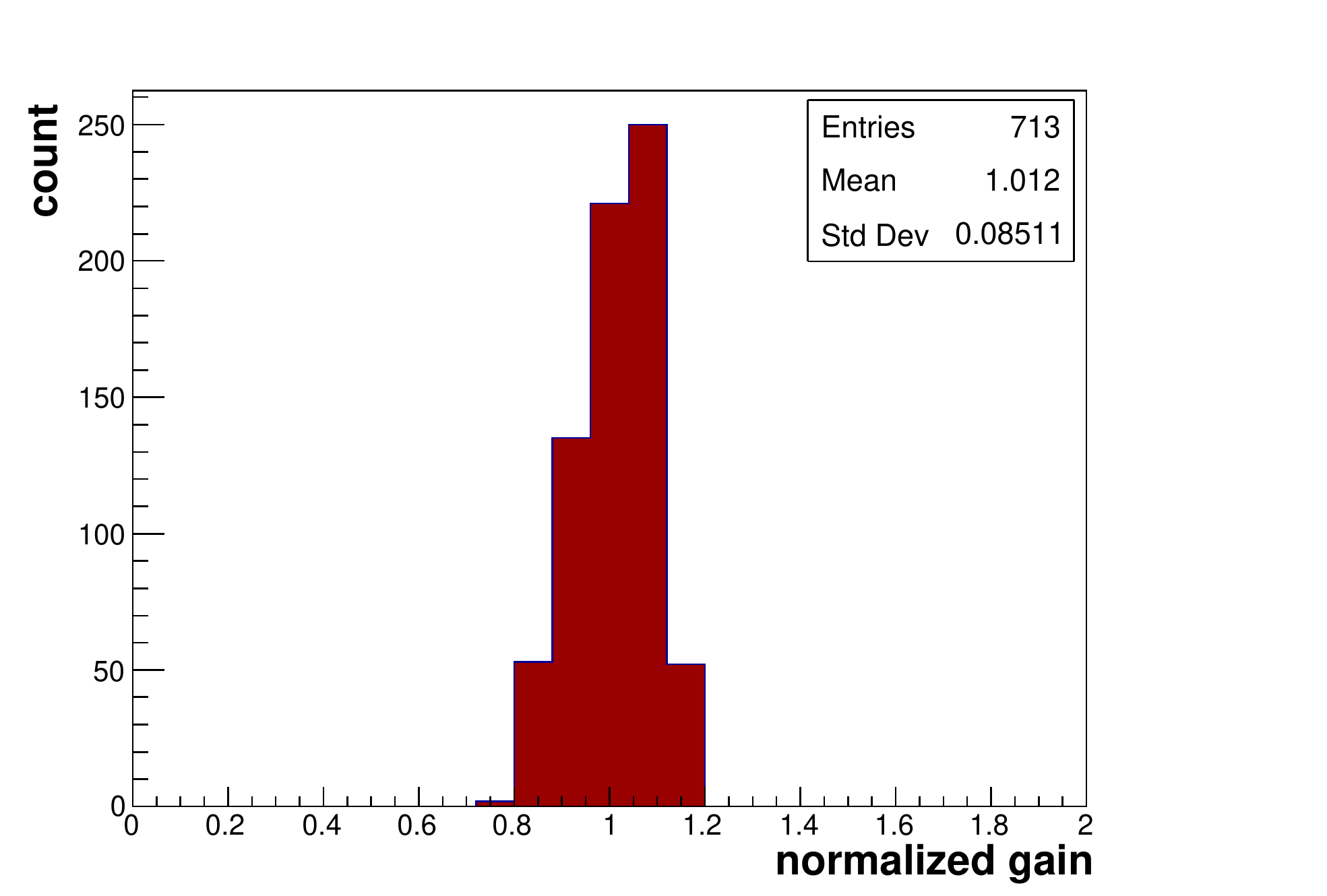}
		\caption{\label{label}Distribution of normalized gain for Ar/CO$_2$ (90:10) gas mixture}
	\end{figure}

		\begin{figure}[htbp]
			\centering
		\includegraphics[scale=0.50]{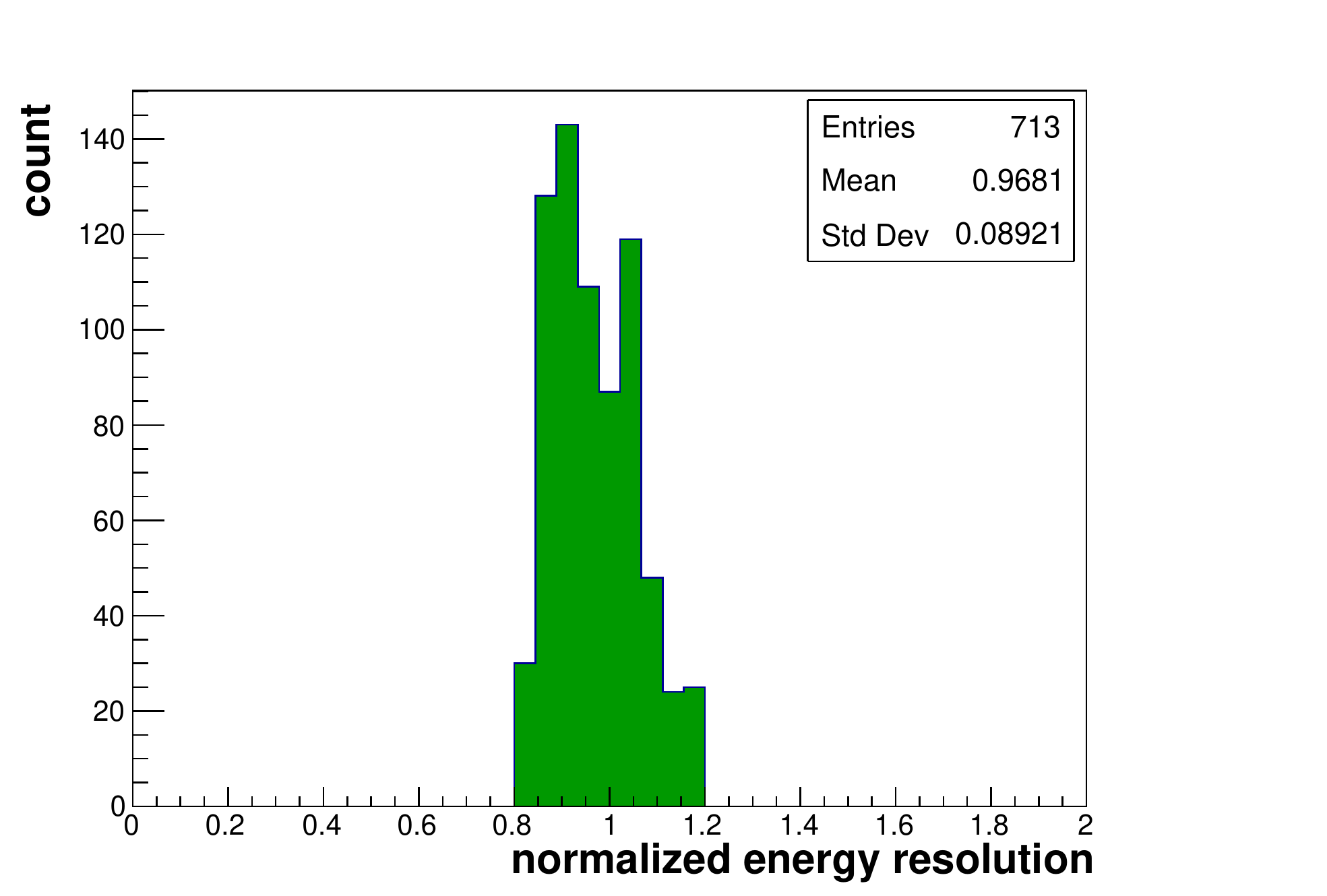}
		\caption{\label{label}Distribution of normalized energy resolution for Ar/CO$_2$ (90:10) gas mixture}
\end{figure}

Figure 1 and 2 show the variation of gain and the energy resolution of the prototype as a function of time along with the ratio of ambient temperature~(T=t+273) and pressure~(p) for (80:20) and (90:10) gas mixtures, respectively. Figure 3 and 4 show the variation of normalized gain and normalized energy resolution as a function of the charge accumulated per unit area for (80:20) and (90:10) gas mixtures, respectively. Figure 5 and 6 show the variation in the normalized gain and energy resolution for the (80:20) gas mixture and figure 7 and 8 show that for the (90:10) gas mixture, respectively.

The variation in the normalized gain and energy resolution has been found to be $\sim$10\% for both the (80:20) and (90:10) gas mixtures as shown in figure 5, 6, 7 and 8.

\section{Conclusions}
A systematic study on the stability of  gain and energy resolution of a prototype triple
GEM detector in long term operation under a high rate of X-ray irradiation has been carried out with Ar/CO$_2$ (80:20) and (90:10) gas mixtures.  For this study, the same Fe$^{55}$ source is used to irradiate the
chamber as well as to measure the gain and energy resolution at an interval of
10 minutes. Using a collimator, the chamber is irradiated with a particle rate of $\sim$20 kHz/mm$^2$ for $\sim$30 hours with Ar/CO$_2$ (80:20) gas mixture and for $\sim$140 hours with Ar/CO$_2$ (90:10) gas mixture which are equivalent to a charge accumulation of 0.6~mC/mm$^{2}$ and 1.8~mC/mm$^{2}$, respectively. No degradation is observed in gain and energy resolution other than a fluctuation of $\sim$10\% after the long exposure to X-ray.

 \section{Acknowledgments}
The authors would like to thank the RD51 collaboration for all kinds of
support. We would like to thank Dr. A. Sharma, Dr. L.
Ropelewski and Dr. E. Oliveri of CERN and Dr. C.
J. Schmidt and Mr. J. Hehner of GSI Detector Laboratory for valuable
discussions and suggestions in the course of the study. This work is
partially supported by the research grant SR/MF/PS-01/2014-BI from
Department of Science and Technology, Govt. of India and the research grant of CBM-MUCH project from BI-IFCC, Department of Science and
Technology, Govt. of India. S. Biswas acknowledges the support of DST-
SERB Ramanujan Fellowship~(D.O. No. SR/S2/RJN-02/2012). Special thanks to Ms. Aayushi Paul for her contribution.

\end{document}